\documentclass{aa}  
\usepackage{graphicx}
\usepackage{natbib}
\bibpunct{(}{)}{;}{a}{}{,}

\usepackage{array}
\usepackage{graphics}
\usepackage{latexsym}
\usepackage{amssymb}
\usepackage{amsmath}
\usepackage{fancyhdr}
\usepackage{float}
\usepackage{morefloats}
\usepackage{slashbox}
\usepackage{multirow}
\usepackage{lscape}
\usepackage[toc,page]{appendix}
\bibpunct{(}{)}{;}{a}{}{,}
\include{hyphe}

\usepackage[english]{babel}
\usepackage{graphicx}

\usepackage{morefloats}
\usepackage{txfonts}
\usepackage{graphicx}
\usepackage{txfonts}
\begin{document} 

   \title{A-PHOT: a new, versatile code for precision aperture photometry} 

   \author{E. Merlin\inst{1}
          \and S. Pilo\inst{1}     
          \and A. Fontana\inst{1}
          \and M. Castellano\inst{1}
          \and D. Paris\inst{1}
          \and V. Roscani\inst{1}
          \and P. Santini\inst{1}
          \and M. Torelli\inst{1}
          }

   \institute{INAF-OAR\\
              \email{emiliano.merlin@oa-roma.inaf.it}
             }

   \date{Received July 31, 2018; accepted November 26, 2018}

  \abstract
   {}
   {We present \textsc{a-phot}, a new publicly available code for performing aperture photometry on astronomical images, that is particularly well suited for multi-band extragalactic surveys.}
   {\textsc{a-phot} estimates the fluxes emitted by astronomical objects within a chosen set of circular or elliptical apertures. Unlike other widely used codes, it runs on predefined lists of detected sources, allowing for repeated measurements on the same list of objects on different images. This can be very useful when forced photometric measurement on a given position is needed. \textsc{a-phot} can also estimate morphological parameters and a local background flux, and compute on-the-fly individual optimized elliptical apertures, in which the signal-to-noise ratio is maximized.} 
   {We check the performance of \textsc{a-phot} on both synthetic and real test datasets: we explore a simulated case of a space-based high-resolution imaging dataset, investigating the input parameter space to optimize the accuracy of the performance, and we exploit the CANDELS GOODS-South data to compare the \textsc{a-phot} measurements with those from the survey legacy catalogs, finding good agreement overall.}
   {\textsc{a-phot} proves to a useful and versatile tool for quickly extracting robust and accurate photometric measurements and basic morphological information of galaxies and stars, with the advantage of allowing for various measurements of fluxes at any chosen position without the need of a full detection run, and for determining the basic morphological features of the sources.} 

   \maketitle
%

\section{Introduction}

Measuring the amount of photons that we receive from astronomical sources is the primary way to gather information about the Universe. Many conceptual and practical subtleties complicate this ideally simple task, such as the precise definition of what should be considered a single object, where its limits projected on the 2D sky should be placed, and how the possible intermission of other sources of emission or absorption on the line of sight should be taken into account (including the terrestrial atmosphere for ground-based observations). However, even considering the ideal situation in which the source is a single, perfectly isolated object, the measurement of its emitted flux from a single scientific photographic image is not necessarily a trivial task.

Ideally, the following steps are required to reach the goal: (i) prepare the image, for instance, by combining different observations into a single mosaic, and correct for flat-field, transients, artifacts, and defects, and so on, (ii) detect the sources above a chosen significance threshold in terms of signal-to-noise ratio (S/N), (iii) separate the sources that overlap one another, for example, by assigning each pixel to a single object (deblending\textup{}), (iv) estimate the extension of each source and its basic morphology, (v) estimate a local background flux, and finally, (vi) measure the flux that is emitted by each object. Currently, no single software performing all these tasks exists, and most of the steps still require some degree of human intervention. In past decades, detection, deblending, and flux measurements have been made efficient and are almost automatized, at least in the extragalactic optical and near- to mid-infrared domain, by the advent of \textsc{SExtractor} \citep{Bertin1996}. 
More recently, Python alternatives have been made available \citep[\textsc{Photutils},][]{Bradley2017}. 


Of course, it is very important to have the possibility to compare the results from different codes rather than having a single choice. In this paper we present \textsc{a-phot}, a new software designed to perform accurate aperture photometry on astronomical images. \textsc{a-phot} assumes that the first three steps of the above list (preparation of the images, detection, and deblending) have been performed, and/or in general, that a list of positions in which the user wishes to measure the fluxes exists: then, \textsc{a-phot} can perform accurate flux measurements within any set of chosen circular or elliptical apertures centered on such positions, as well as local background estimates (steps (iv) to (vi) in the above list). Furthermore, if a \textit{\textup{segmentation}} map exists in which the significant pixels 
belonging to each detected source are univocally identified, \textsc{a-phot} can estimate the basic morphological properties of each source.
This approach has the advantage of giving the full control over the process of flux measurements, detaching it from other duties and making it more transparent, and allowing for the repetition of several measurements on the same list of detected objects or in any position of interest. This feature can be of crucial importance when a measurement of the flux within any given region of a field is needed, whether it is centered on a detected source or not (for any scientific reason). 

Furthermore, \textsc{a-phot} provides an option that allows determining the region around the chosen position in which the S/N is maximized; this can be a powerful way to tackle the disturbance of background noise, for example to estimate colors with great accuracy.

While it is necessary in many cases to adopt more sophisticated methods to accurately measure fluxes, for instance using template fitting to cope with blending in low-resolution images \citep[with codes like \textsc{t-phot}, see][]{Merlin2015,Merlin2016b}, plain aperture photometry is still the reference technique when possible because it is much faster and straightforward in its approach. However, it must be pointed out that when two sources are close enough to overlap, any aperture measurement will be affected, despite attempts to mitigate the issue by removing or replacing contaminated pixels, for instance. In general, this is not an issue when working on high-resolution images, such as Hubble optical or near-infrared images, but it can become a crucial problem at redder wavelengths and/or poorer seeing. 
Combining the two techniques can enhance the robustness of the results by lowering the uncertainties related to a single method. 

The paper is organized as follows: in Section \ref{code} we describe the code and its main algorithms, and in Section \ref{tests} we present and discuss a set of tests we performed to assess its performance. Finally, in Section \ref{conclusions} we briefly summarize the results. The code is publicly available for download at the website \textit{http://www.astrodeep.eu/a-phot/}.

\section{Description of the code and basic algorithms} \label{code}

\textsc{a-phot} is written in \texttt{C}, with a Python wrapper. In the tarball, the user will find a documentation file in which the technical specifications about installation and usage are given. Here we briefly summarize the code features.

At variance with \textsc{SExtractor}, \textsc{a-phot} is designed to perform photometric measurements requiring a detection catalog (and optionally a segmentation map) \emph{\textup{as an input}}. Nevertheless, it is also possible to internally compute some morphological parameters (semi-axis lengths, ellipticity, position angle, and Kron radius), and to compute and subtract a local background. These options are described in Sections \ref{morpho} and \ref{bkg}.

\subsection{Input}

\textsc{a-phot} requires the following input:
\begin{itemize}
\item A scientific image (SCI) in \texttt{FITS} format.
\item The corresponding root mean square (RMS) map, giving the flux uncertainty in each pixel.
\item A catalog listing the detected and deblended sources in a fixed format:
\texttt{ID x y R e theta}, where \texttt{ID} is the source identification number, \texttt{x} and \texttt{y} are its position in pixel space or in WCS (see below), \texttt{R} is the major semi-axis of the elliptical aperture on which the measurement will be performed, \texttt{e} is the ellipticity ($e=1-b/a$ where $a$ and $b$ are the semi-axes of the ellipse) and \texttt{theta} is the position angle. \texttt{R}, \texttt{e,} and \texttt{theta} can also be estimated internally (see Section \ref{morpho}). 
A Python script to automatically build the catalogue in the correct format from a standard \textsc{SExtractor} detection catalog is provided.
\item Optionally, a segmentation map and a flag map can be given (see below). We note that it is mandatory to input the segmentation map to perform the internal estimation of morphological parameters.
\end{itemize}

\begin{table}
\caption{Photometric flags reported in the output catalog. The final flag is the sum of the individual flagging values.} \label{flags}
\centering
\begin{tabular}{|p{1.5cm}|p{6.5cm}|}     
\hline
Flag & Description \\ \hline\hline
+0 & Regular source \\ \hline
+1 & Source is contaminated by close neighbors, or has bad pixels \\ \hline
+2 & Source is blended with another  \\ \hline
+4 & Source is saturated \\ \hline
+8 & Source is close to a border \\ \hline
\end{tabular}
\end{table}

A configuration file must be fed to the code upon launching a run; it must specify the  above inputs, along with some other parameters and options, among which the lists of circular diameters and elliptical semi-axes to construct the apertures within which the flux computation is desired (see Section \ref{flux}); these can have any value, so that they are not dependent on the structural parameters given in the input catalog. A template configuration file is provided with the code and can be printed by launching the code with the command line option \texttt{-p}. Parameters can also be specified by command line, overwriting those given in the configuration file, as described in the documentation included in the code tarball. 

When the input coordinates are given in pixel space, then \texttt{R} in the input catalog and the apertures in the configuration file must also be in pixels, while \texttt{theta} must be given in degrees, counter-clockwise with respect to the image $X$ axis. Conversely, if the positions are given in WCS (as Right Ascension and Declination), then \texttt{R} and the apertures must be in arcseconds, and \texttt{theta} (again in degrees) must be the position angle with respect to the equatorial plane. Finally, \texttt{e} is a dimensionless parameter. The choice between pixel space and WCS input must be specified in the configuration file by means of a dedicated keyword.

\subsection{Output}

The output consists of a catalogue in text format, which lists
\begin{itemize}
\item ID;
\item $x$ and $y$ pixel coordinates;
\item the fluxes computed within the desired circular apertures (in the same units as the scientific image, or if a dedicated flag is switched on and a photometric zero-point is provided, directly in $\mu$Jy);
\item the fluxes computed within the desired elliptical apertures;
\item the uncertainties corresponding to the circular apertures;
\item the uncertainties corresponding to the elliptical apertures;
\item the value of the brightest pixel, in counts/s;
\item a quality flag assigned to the object, see Table \ref{flags};
\item if required, the value of the measured background \emph{\textup{per pixel}}.
\end{itemize}

A companion catalog can be produced at the same time, listing magnitudes instead of fluxes (a photometric zero-point $ZP$ must be given as input; magnitudes are computed straightforwardly as $m=-2.5\times log_{10}F+ZP$, where $F$ is the flux of the source within the considered aperture), plus the value of $\mu_{max}$, that is, the surface brightness of the brightest pixel of each object (in mag/arcsec$^2$).
If morphological parameters are computed internally, their values are also written in a separated catalog.


\subsection{Flux measurement and uncertainties} \label{flux}

\begin{figure}[t!] 
  \includegraphics[width=9cm]{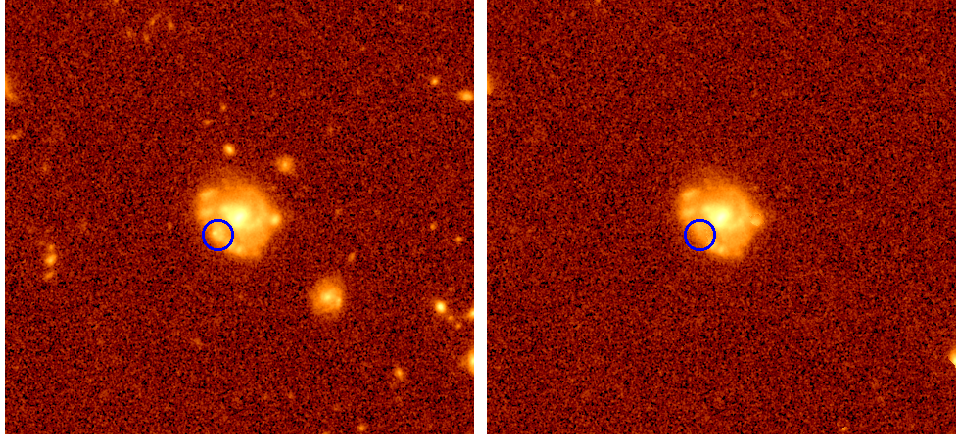}
  \centering
\caption{Effects of the symmetric pixel-replacement option. Left: original image; right: image where pixels belonging to close contaminating objects have been replaced with the symmetric pixels with respect to the center of the source being analyzed. The blob in the blue circle, which had been (erroneously) classified as a separate source in the detection process, has been replaced as well.}\label{repl}
\end{figure} 

To obtain the fluxes within a given (circular or elliptical) aperture, \textsc{a-phot} first sums the values of the pixels that completely fall inside the aperture. 
If a segmentation map is provided as input, the values of the pixels within the aperture that are assigned to another object can either be replaced with the values of the symmetric pixels with respect to the centroid of the source (the effects of this choice are shown in Figure \ref{repl})\footnote{When the source is close to an edge of the image, it may be impossible to perform such a centrally symmetric substitution because the target pixels might be outside the matrix of pixels. In this case, an axially symmetric substitution is performed instead when possible. If this is not possible either, the pixel is flagged out and excluded from the summations.}, or simply not be included in the summation (a dedicated keyword in the configuration file is used to choose between the two options). The same holds for pixels that are flagged as bad if a flag map is provided, and for those with RMS values above a given threshold.

The pixels that are crossed by the border of the considered circular or elliptical aperture are divided into a grid of $n \times n$ sub-pixels (where $n$ is the value given in the configuration file) and the flux from the sub-pixels inside the aperture is added to the summation.\footnote{A pixel $x,y$ is considered to be within an elliptical aperture centered on $x_0,y_0$ if
\begin{equation}
  x_1^2/a^2 + y_1^2/b^2 < 1,
\end{equation}

\noindent where $x_1=(x-x_0) \times \mbox{cos$\theta$} + (y-y_0) \times \mbox{sin$\theta$}$ and $y_1=-(x-x_0) \times \mbox{sin$\theta$} + (y-y_0) \times \mbox{cos$\theta$}$, and $\theta$ is the aperture position angle.}

Uncertainties are straightforwardly computed starting from the formula known as the "CCD Equation" \citep{Mortara1981},

\begin{equation}
  \frac{S}{N} = \frac{f}{\sqrt{f+n_{pix}(N_S+N_D+N_R^2)}},
\end{equation}

\noindent where $f$ is the total amount of photons received, $N_S$ is the number of photons per pixel from the background, $N_D$ is the number of dark current electrons per pixel, and $N_R$ is the number of read-out noise electrons per pixel. 
When we assume that the RMS map accounts for all the three terms in brackets, this translates into a summation of the squared values of the RMS pixels corresponding to those included in the flux computation, plus the term accounting for the photon shot-noise: 
\begin{equation}
  \sigma_{obj}=\sqrt{\sum_{i=1}^{N_{pixels}} RMS_{i,obj}^2 + f_{obj}/G}.
\end{equation}

\noindent Here $G$ is the \textit{\textup{gain}}, that is, a factor that accounts for the efficiency of the CCD in transforming the signal from photons to electrons, and it is assumed that everything has been scaled to 1 sec of exposure\footnote{If the RMS map accounts for this term as well, the second term is not necessary and it can be excluded by setting \texttt{GAIN=0} in the configuration file.}.

\textsc{a-phot}  measures the flux within any chosen number of circular and elliptical apertures by looping once on the pixels where the object is found and summing separately the contributions for each of the desired apertures; this technique ensures an efficient and time-saving performance. 
The circular apertures must be given in the configuration file as the diameter (in pixels) of the desired circle to be centered on the desired positions. The elliptical apertures must be input as the multiplicative factors to be applied to the value of $R$ of the sources given in the input catalog. When the aim is to estimate the total flux of the object, $R$ can be taken to be some multiple of the Kron radius $R_{Kron}$ \citep{Kron1980}, which in turn can also be determined internally (see next section); a further possibility is to compute the flux within a "maximum S/N" elliptical aperture, computed on-the-fly object by object (see Sect. \ref{optsn}). It is worth pointing out that the possibility of having multiple measurements on geometrically similar elliptical apertures, and in particular to compute an "S/N maximizing aperture", is not present in similar codes for aperture photometry such as \textsc{SExtractor}. 

Since in some unfortunate cases the flux fluctuations around objects could include unphysically negative spots, a "clipping" option is included. When switched on, only pixels with values above $-min(\sigma_{RMS},\sigma_{bkgd})$ are included in the summation (where $\sigma_{RMS}$ is the value of the pixel in the RMS map, and $\sigma_{bkgd}$ is the standard deviation of the distribution of pixel values in the region where the background is computed, see Section \ref{bkg}; note that the two options "clipping" and "background subtraction" are independent from one another, i.e. they can be switched on and off separately).


\subsection{Estimation of morphological parameters} \label{morpho}

\textsc{a-phot} can compute some basic morphological parameters of the detected objects: namely, the lengths of the semi-axis $a$ and $b$ of the isophotal ellipse that approximates the shape of the source (and hence the ellipticity and/or elongation), and the position angle (the inclination of the major semi-axis with respect to the $X$-axis of the image, measured counter-clockwise). The latter can be determined after computing the second moments of the light distribution, summing on all the significative pixels, that is, those belonging to the segmented area of the sources. The segmentation map is mandatory to accomplish this task. 

The first and second moments of the light distribution are computed as
\begin{eqnarray*}
  \bar{x} &=& \sum f x / \sum f  \\
  \bar{y} &=& \sum f y / \sum f  \\  
  \bar{x^2}&=&\sum f x^2 / \sum f - \bar{x}^2  \\
  \bar{y^2}&=&\sum f y^2 / \sum f - \bar{y}^2  \\
  \bar{xy}&=&\sum f x y / \sum f - \bar{x}\bar{y,}
\end{eqnarray*}

\noindent where $f$ is the value of the flux of each pixel, and the summation is performed including the pixels within a circular region centered on the input coordinates of the centroid of each source, having a radius estimated to include the whole segmented area.

From these equations, the values of the semi-axes $a,b$ and of the position angle $\theta$ can be obtained as

\begin{eqnarray}
  a&=&\sqrt{(\bar{x^2}+\bar{y^2})/2 + \sqrt{[(\bar{x^2}-\bar{y^2})/2]^2+\bar{xy}^2}}; \\
  b&=&\sqrt{(\bar{x^2}+\bar{y^2})/2 - \sqrt{[(\bar{x^2}-\bar{y^2})/2]^2+\bar{xy}^2}}; \\
  \mbox{tan}(2\theta)&=&2\bar{xy}/(\bar{x^2}+\bar{y^2}).
\end{eqnarray}

These parameters can also be given as external input, for example, taking them from the same \textsc{SExtractor} catalog from which the positions and segmentation have been obtained. Comparisons between the internally computed values and those taken by \textsc{SExtractor} show good overall consistency. 

The Kron radius $R_{Kron}$ of the object within which $\sim 90\%$ of the total emitted light is expected to fall \citep[see, e.g.,][]{Graham2005}, is finally obtained by summing on the pixels within an extended circular region (typically of radius $6 \times a$), as
\begin{equation}
  R_{Kron} = \sum r f / \sum f
,\end{equation}

where $r$ is the distance of each pixel to the centroid of the object. The elliptical "total" aperture fluxes (corresponding to \textsc{SExtractor} \texttt{FLUX\_AUTO}) are then computed y summing within the elliptical area with major semi-axis $R_{Kron}' = 2.5 \times R_{Kron}$, in an attempt to recollect a fraction closer to 100\% of the real total flux (around 94\% following the \textsc{SExtractor} manual; see Section \ref{flux}).

When the background subtraction option is switched on, the pixels values must be considered \emph{\textup{after}} this subtraction. Therefore, if the morphological parameters are computed internally, a double loop is performed, following this conceptual scheme: (i) a first guess of the Kron radius is obtained from the original image (i.e., before any local background subtraction); (ii) this radius is used to select the region in which a first background estimation is performed; and (iii) a second estimate of the Kron radius is then made, using the background subtracted values of the pixels. Finally, (iv) a new background subtraction is made using this Kron radius estimate (see Section \ref{bkg}).

\subsubsection{Minimum size of the Kron aperture} \label{kmin}

It is common practice to impose a minimum and a maximum value to $R_{Kron}'$ (i.e., the ``enlarged'' area with semi-axis $2.5 \times R_{Kron}$ in which the ``total'' flux of the source is computed). Despite the fact that these thresholds typically have standard default values (in units of the only available scale length, i.e., the major semi-axis of the isophotal aperture $a$), it turns out that the choice of the minimum threshold $R_{Kron,min}'$ has a strong impact on the accuracy of the results. 
One should keep in mind that the concept of ``Kron radius'' has been elaborated some decades ago, in the early 1980s, 
to be applied on local galaxies with low S/N observed from ground-based facilities. Since then, the requirements on photometric accuracy have become more and more demanding, which means that just straightforwardly applying the original recipes to space-based observations of high-redshift sources, with a wide range of S/N ratios but often with structural and observational features much different from local galaxies, might not be an appropriate choice:
not only is a robust computation of the light moments of such sources difficult and often marginally reliable, but furthermore, the small number of significant pixels generally causes a collapse of the computed Kron radii toward very low values. As an example, in the \citet{Guo2013} CANDELS catalog of the GOODS-South field, in which 34930 objects detected in the $H$-band with \textsc{SExtractor} are listed, $\sim 27\%$  have $R_{Kron}'$ equal to the minimum allowed value of $3.5 \times a$. 

To better understand the issue, we ran a test to measure the fraction of the total flux within self-similar elliptical apertures of increasing size. We simulated $\sim12000$ galaxies by means of the software \textsc{Egg} \citep{Schreiber2016}, which creates a mock cosmological catalog based on the observed properties of the CANDELS survey \citep{Grogin2011}. The galaxies in the catalog are double-Sersic models with realistic bulge and disk spectral energy distributions (SEDs). We then transformed the catalog into FITS stamps mimicking an optical survey from space (limiting magnitude $m_{lim}=27.0$ at 10$\sigma$, FWHM=0.2", pixel scale = 0.1"), using the \textsc{SkyMaker} code \citep{Bertin2009}. 
For each image, a point spread function (PSF) was generated automatically by the code, taking into account the desired specifications. On each stamp, containing a single source and large enough to include its potentially extended faint wings, we ran \textsc{a-phot} to estimate the morphological parameters $a$, $e,$ and $\theta$. Then, we iteratively measured the flux within a set of elliptical apertures centered on the galaxy, with identical shape but varying size.

\begin{figure}[t!] 
  \includegraphics[width=9cm]{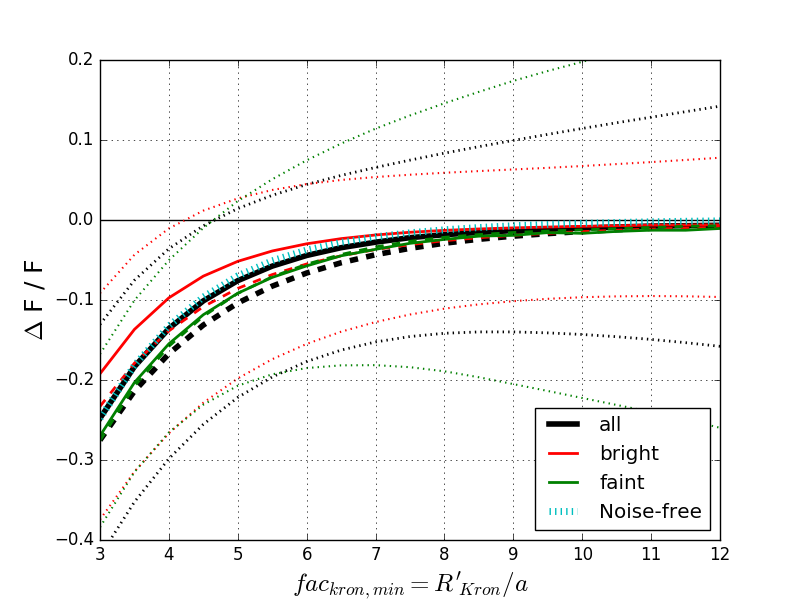}
  \centering
\caption{Lost fractions $\Delta F / F$ of the total true flux of $\sim$12000 simulated galactic objects. The fluxes have been measured within elliptical apertures whose major semi-axis is equal to a given multiple $fac_{kron,min}$ of the object isophotal semi-axis $a$. The solid lines show the median, the dashed lines the mean, and the dotted lines the standard deviations. The black line represents all sources, the red line only the bright ($m$<24.5) sources, and the green line only the faint (25.5<$m$<26.5) sources. The dotted cyan curve corresponds to a full run on images where the observational noise was not included. Imposing $R_{Kron,min}'=3.5\times a $, as typically done, yields a substantial underestimation of the flux. See text for more details.}\label{kronmin}
\end{figure}

The result of the test is the growth curve plotted in Fig. \ref{kronmin}, showing the missing fraction of the flux $\Delta F / F$ (zero means that the totality of the flux has been recovered) as a function of the multiplicative factor $fac_{kron,min}$ applied to the object estimated semi-major axis $a$ to obtain the actual dimensions of the elliptical aperture; in other words, $fac_{kron,min}$ is the dimension of the actual elliptical aperture in units of the semi-major isophotal axis $a$, $fac_{kron,min}=R'_{Kron}/a$.
Black lines correspond to the full simulated sample, while green and red lines refer to only faint (25.5<$m$<26.5) and only bright ($m$<24.5) objects; the solid lines are the medians of the $\sim12000$ measurements, the dashed lines are the means, and the dotted lines are the standard deviations from the means (the simulated images contain Gaussian observational noise, which  means that individual measurements oscillate around the average values). 
As expected, the missing fractions tend to zero with larger apertures; the bright objects converge faster than the faint ones. However, imposing a minimum radius $R_{Kron,min}'= 3.5 \times a$ (a typical choice) can lead to a severe underestimation of the flux of the objects, on average,  $\sim 15-20\%$ of the photons are lost. To recover more than 95\% of the flux, for instance, $fac_{kron,min}$ should be at least $\sim 7-8$, at least in this test\footnote{In the CANDELS GOODS-South catalog, $\sim 97\%$ of the sources have $R'_{Kron}<8.0 \times a$, meaning that most of the detected galaxies might have an underestimated total flux.}.

The cyan dotted line corresponds to a run on the full sample in which the images were simulated without the observational noise: this ensures that the results are \textit{\textup{on average}} close to the ideal case in which the noise does not affect the measurements. Clearly in the real cases, in contrast, the individual measurements are severely affected by the noise (as the thin dotted lines show); in particular, the scatter begins to diverge for faint sources when $fac_{kron,min} \geq 7$. 

Of course, we recall that this is just a single case, a particular realization, and should not be generalized to all the possible situations. Nevertheless, the global picture is clear: enlarging the measurement to larger radii is necessary to statistically retrieve a large portion of the total flux of the sources, provided that contamination effects are correctly taken into account.
In Section \ref{tests} we show that this choice yields good results in the typical situation of extragalactic surveys. In general, for each dataset, the best practice would be to find the most effective values of the parameters by means of an independent set of simulations.
Clearly, this approach might worsen the accuracy of the measurements when contamination issues are considered as well; however, for high-resolution images, this can be considered a minor problem, provided the flux from pixels assigned to nearby sources is excluded from the summation. 

\subsubsection{Estimation of optimal S/N aperture} \label{optsn}

\begin{figure}[t!] 
  \includegraphics[width=8cm]{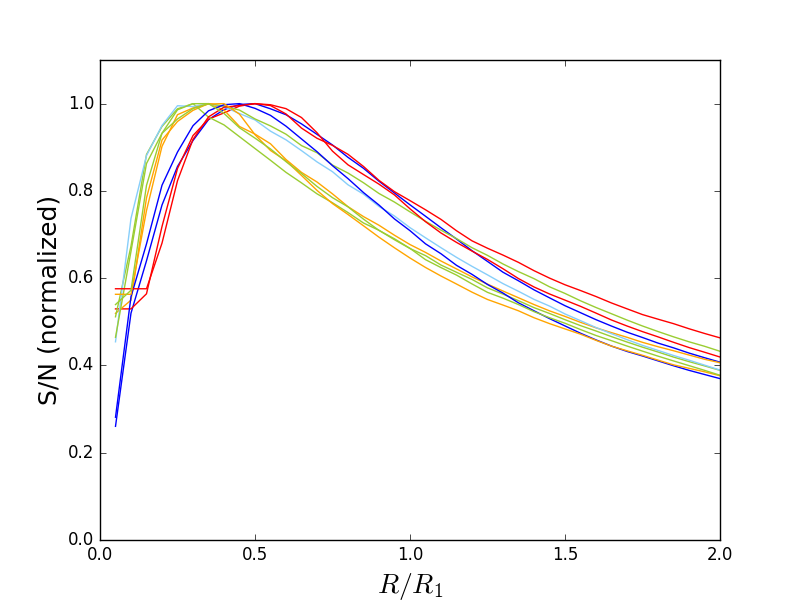}
  \includegraphics[width=7cm]{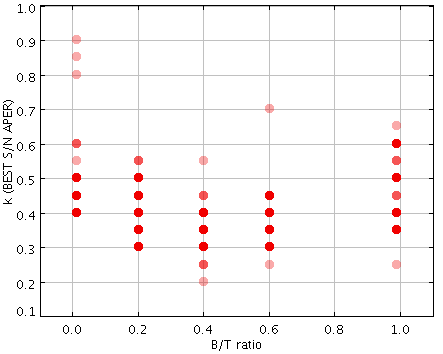}
  \centering
\caption{Top: Typical trends of S/N values within apertures with major semi-axis $R$, as a function of the normalized $R/R_{Kron}'$, for a subset of simulated double-Sersic galaxies. The curves are color-coded to show the morphological type of the synthetic galaxies, with red representing bulge-dominated objects, blue representing disks, and yellow-green representing double components objects with different bulge-to-total ratios. Bottom: values of the best S/N aperture dimensions as a function of the bulge-to-total ratio.}\label{sn}
\end{figure}

\begin{figure*}[t!] 
  \includegraphics[width=14cm]{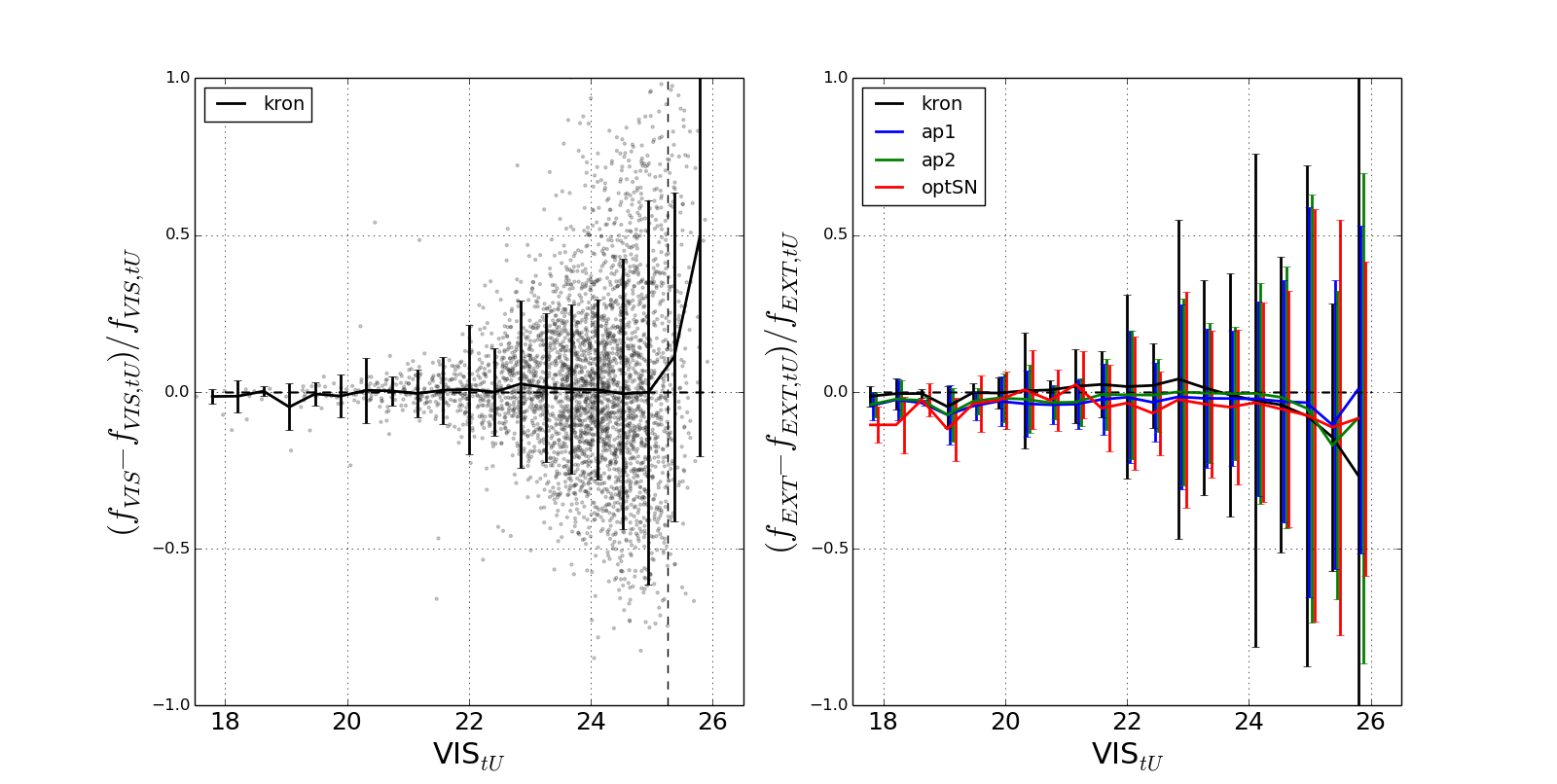}
  \includegraphics[width=14cm]{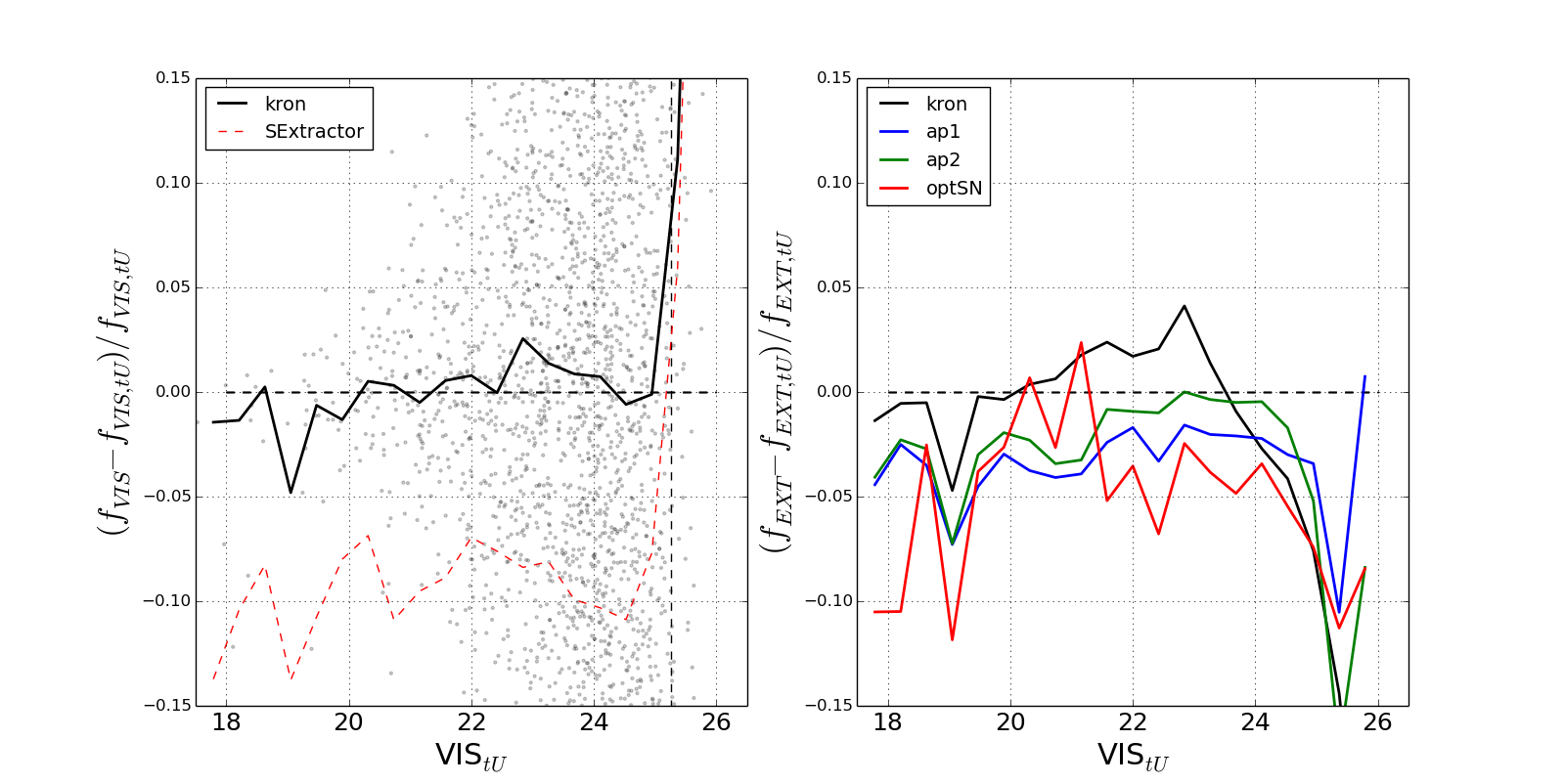}       
  \centering
\caption{Accuracy in flux determination on simulated datasets. Upper panels: full dataset; lower panels: zoom in the central strips; left panels: HRI flux determination; right panels: LRI flux determinations, using different methods. All plots show the quantity $(f_{meas}-f_{input})/f_{input}$ versus input HRI magnitude: the median offset is consistent with zero in the HRI case, and typically below 5\% in the LRI cases. The red dashed line in the bottom left panel refers to a \textsc{SExtractor} run. See text for details.}\label{euc}
\end{figure*}

\begin{figure*}[t!] 
  \includegraphics[width=12cm]{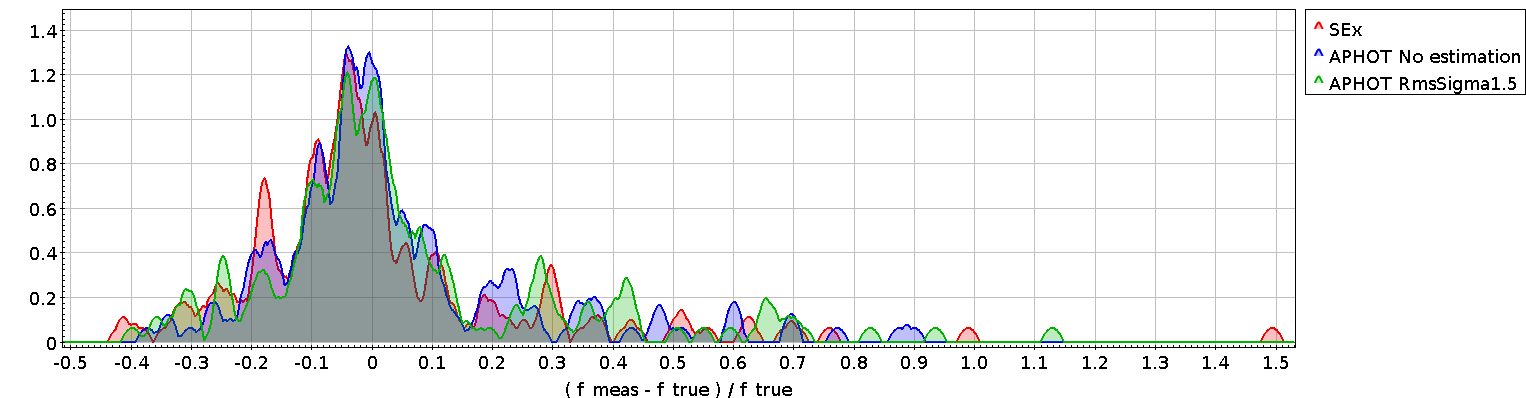}
  \includegraphics[width=12cm]{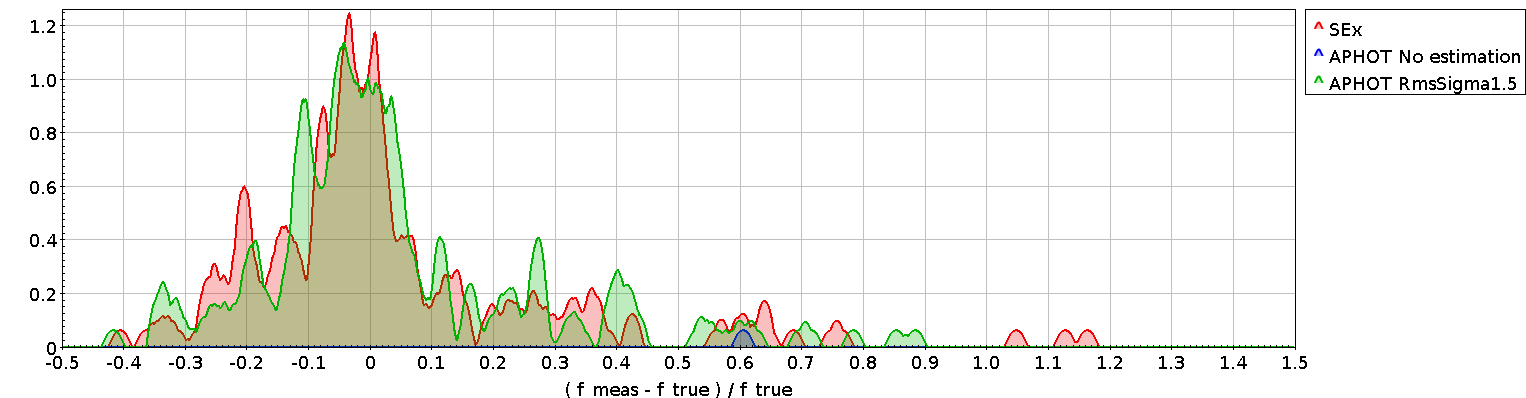}
  \centering
\caption{Compared accuracy of background estimation. Top panel: Histograms of the relative error in flux determination on a simulated image in which no background is present; \textsc{SExtractor} (which automatically computes and subtracts a local background), \textsc{a-phot} including background measurement and \textsc{a-phot} without background estimation yield very similar results. Bottom panel: Same for a simulated image in which a constant background plus local variations in the background intensity have been added; in this case, \textsc{a-phot} without background estimate yields incorrect results (out of scale in this plot), while \textsc{SExtractor} and \textsc{a-phot} with background estimation continue to yield similar results.
}\label{bkgf}
\end{figure*}

\begin{figure*}[t!] 
  \includegraphics[width=16cm]{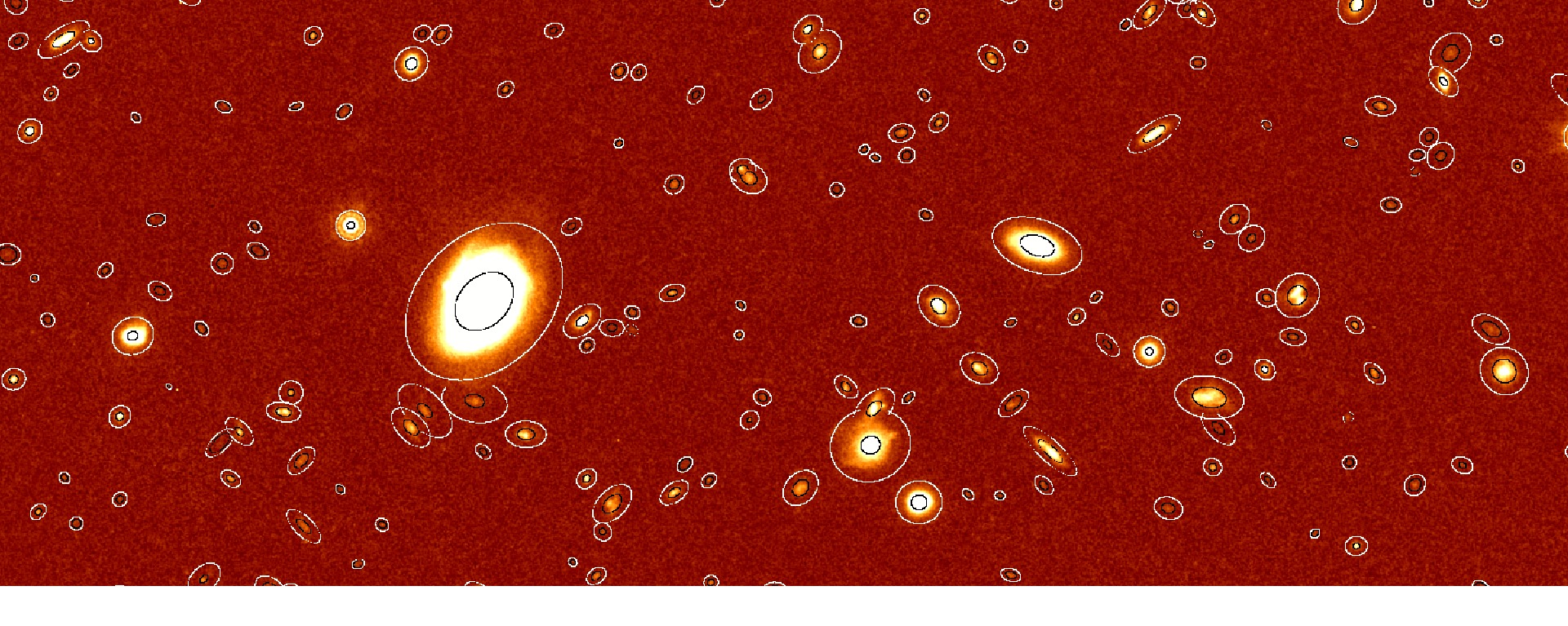}  
  \centering
\caption{Portion of the GOODS-South WFC3 $H160$ CANDELS mosaic. 
White ellipses show the Kron apertures, and black lines the corresponding apertures maximizing the S/N; both have been computed internally by \textsc{a-phot}.}\label{apertures}
\end{figure*}

\begin{figure}[t!] 
  \includegraphics[width=8cm]{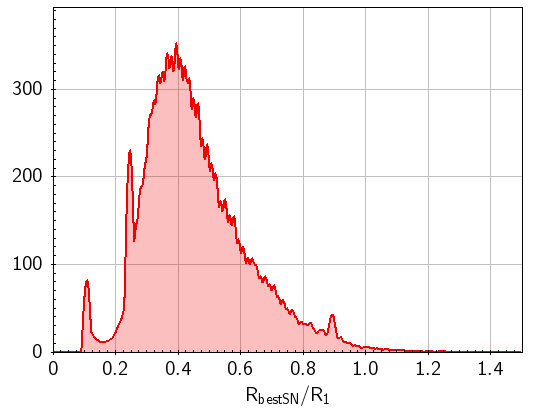}
  \centering
\caption{Ratio between the computed semi-axis of the apertures optimizing S/N and the Kron aperture in GOODS-South $H160$.}\label{histGS}
\end{figure}

\begin{figure}[t!] 
  \includegraphics[width=9cm]{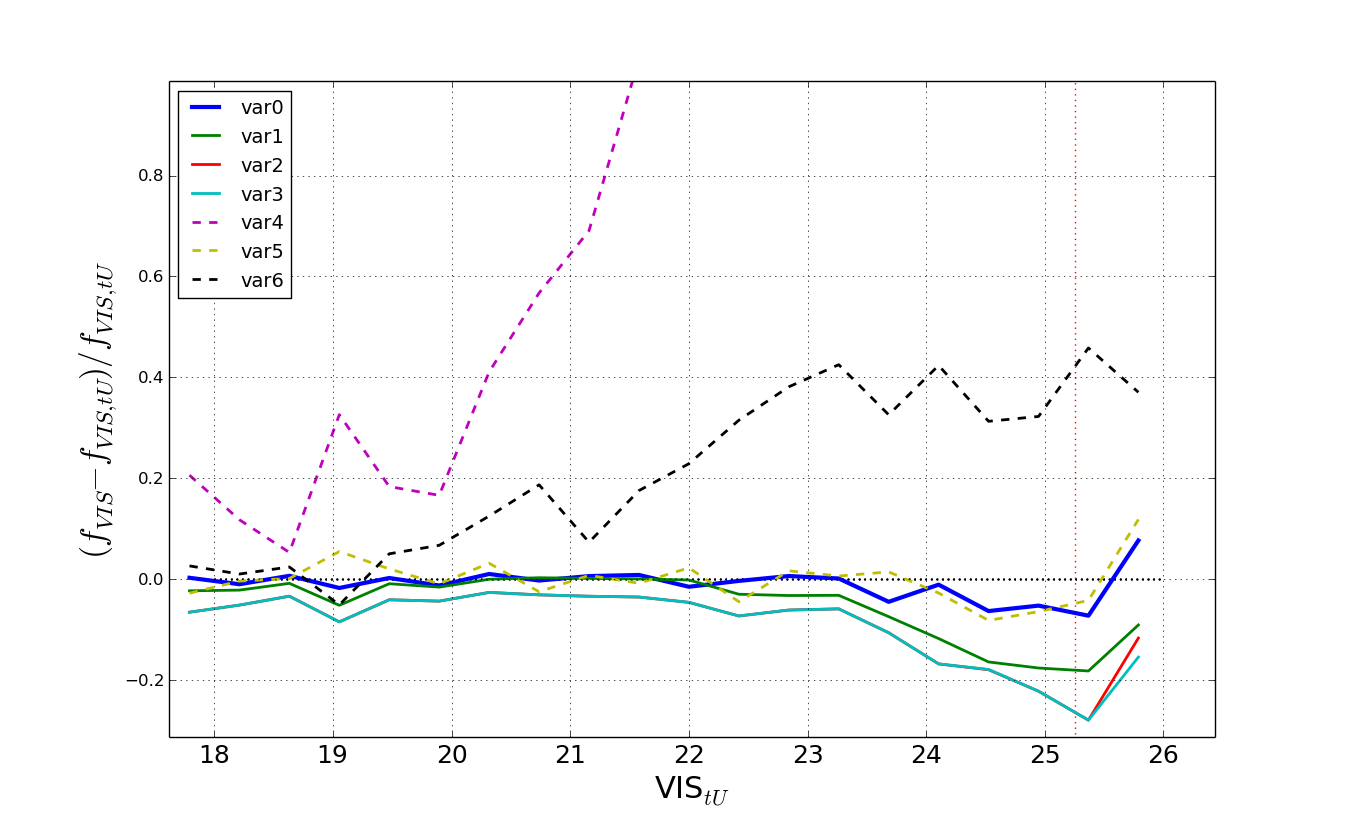}
  \centering
\caption{Compared accuracy of \textsc{a-phot} on a simulated Euclid VIS-like image using the different input parameters listed in Table \ref{tab1}. The plots show the relative difference between the measured  and the input flux as a function of input magnitude. Solid lines correspond to runs on the normal simulated image, whereas dashed lines refer to runs in which an artificial constant background has been added to the image. See text for details.}\label{params}
\end{figure}

A particularly useful option of \textsc{a-phot} is the automatic computation of the dimensions of an elliptical aperture within which the total S/N is maximized. To do so, \textsc{a-phot} uses a recursive bisection algorithm, measuring the fluxes and uncertainties within a set of elliptical apertures, and iteratively getting closer to the one that maximizes the S/N, until convergence below a given tolerance threshold is reached. This feature can be of interest when the impact of background noise for a comparison of the fluxes in two bands needs to be reduced, since measuring the color in the region of PSF-matched images of two sources where the S/N is maximized can reduce the uncertainties and allow for better photo-$z$ estimates.


It is interesting to check whether some systematic trend exist, at least for regularly shaped objects (ellipticals and disks). Using the simulated stamps of galaxies described above, we measured the S/N values within a number of similar elliptical apertures of increasing dimensions. We found as predictable that the S/N follows a typical trend with the aperture radius, increasing monotonically up to a maximum, and then decreasing subsequently at larger radii. It is interesting to note that in the considered case, the maximum tends to be found close to half the value of the measured Kron radius $R_{Kron}'$ (here computed using the typical minimum factor, $R_{Kron,min}'$=3.5). The top panel of Fig. \ref{sn} shows this result: it depicts the (normalized) value of the S/N measured within a normalized ellipse with major semi-axis $R/R1$ as a function of the same radius for a well-behaved selection of simulated objects. The colors of the lines in the plot are a proxy for the input morphological types of the galaxies, with blue being pure disks, red being pure bulges, and greenish being objects including a bulge and a disk with variable bulge-to-total ratios (B/T). The semi-major axis of the ellipse within which the S/N is maximized is typically in the range $0.3 < R/R_{Kron}' < 0.6$, with a mild dependence on the morphology of the object: pure disks and pure bulges seem to have slightly higher values than objects including a disk and a bulge (as shown in the bottom panel), a possible explanation being that the Kron radius is estimated on the disk extension, but the concentration of light in the bulge tends to shrink the S/N radius. It must be pointed out that outliers to this trend do exist, although they represent a negligible fraction of the objects; they usually are the results of inaccurate Kron radius estimations, typically for elliptical galaxies with a large Sersic index, whose true "limits" are very hard to define because of the extended faint wings of the light distribution.

Dealing with real data is obviously more complicated, as we do not know the real value of the bulge/disk decomposition as in synthetic images. The global behavior is similar, however (see Section \ref{GS} for an applicative example).

\subsection{Estimation of local background} \label{bkg}

\textsc{a-phot} can compute local background estimations, writing background-subtracted fluxes in the output catalog. To this aim, \textsc{a-phot} computes the mean value of the pixels (i) outside a radius $r_{bkgd}=fac_{rbkgd} \times R_{Kron}'$, (ii) within $r_{bkgd}+r_{buf}$ pixels, excluding (iii) those that belong to the considered object according to the segmentation map (if given), and (iv) those with a flux value $f$ so that $|f-f_{median}| < q_1 \times \sigma_{RMS}$, where $fac_{rbkgd}$ and $r_{buf}$ are free parameters to be given in input and $\sigma_{RMS}$ is the value of the pixel in the RMS map. Furthermore, this mean is recursively recomputed with a clipping procedure, excluding the pixels that deviate from the median for more than $q_2 \sigma_{bkgd}$ (as a caveat, we note that this procedure is likely to bias - underestimate - the background in the Poisson regime, e.g., when the background is low, for small pixels and negligible read-out noise). See Section \ref{tests} for a discussion of the typical values of all these parameters.

If the number of pixels gathered in this way is below a threshold $n_{min} = 0.25 \times A$ (where $A$ is the area of the ring in which the background is evaluated), the value of $q_2$ is iteratively increased until the threshold is reached or a maximum number of iterations is done.

Following the approach implemented in \textsc{SExtractor}, if the final value of the median after these iterations is varied by less than 20\% with respect to the initial value, it is assumed that the background is fairly flat and the mean value of the clipped pixel is taken as the background flux; otherwise, the expression $2.5 \times f_{median} - 1.5 \times f_{mean}$ is used.

\section{Tests} \label{tests}

In this section we check the performance of \textsc{a-phot} by means of a number of dedicated tests. The typical computational times are comparable to the corresponding \textsc{SExtractor} runs.

%
%

\subsection{Measurements on simulated images} \label{sims}

To test the global performance of \textsc{a-phot}, we used the mock cosmological catalog created using \textsc{Egg}, and we produced a full simulated field of view, with a high-resolution image (HRI in the following) mimicking an Euclid-like VIS coadd, and a low-resolution image (LRI) mimicking a ground-based $g$ coadd, taking from the Euclid Red-book \citep{Laureijs2011} the expected values of depths, zero-point, and seeing (FWHM=0.2", limiting magnitude 27 AB at 1$\sigma$ for HRI, and FWHM=0.8", limiting magnitude 27.7 AB at 1$\sigma$ for LRI). Again, we used the PSFs built by \textsc{SkyMaker} to produce the images. 
On both images we performed a detection and deblending run using \textsc{SExtractor} to obtain the input positions and segmentation maps. Then, we used \textsc{a-phot} to estimate the morphological parameters, and we then measured the fluxes of the sources, counter-checking the output with the input "true universe" values after cross-correlating on a positional basis. 

Figure \ref{euc} shows the overall accuracy of \textsc{a-phot} in the estimation of the total fluxes of the galaxies, using different methods that combine the output of the code. The lower panels are magnifications of the central strip of the upper ones. The left panels show the average relative error in the HRI measured total flux, $df=(f_{meas}-f_{input})/f_{input}$, computed on the simulated image using the internally computed elliptical Kron radius. The median of $df$ is consistent with zero, and with no measurable offset, down to the detection limit. For comparison, the dashed red line is the $df$ obtained taking the \textsc{SExtractor} \texttt{FLUX\_AUTO} (which corresponds to the Kron flux, as obtained in the detection run) value as $f_{meas}$; noticeably, $\sim10\%$ of the flux is lost using this straightforward approach. This can be understood since, as already pointed out, in typical cases, $\sim 1/3$ of the detected sources have a Kron aperture equal to the minimum allowed value $3.5 \times a$, which as shown in Section \ref{kmin} can severely underestimate the total flux of the objects, and $\sim 80\%$ are below the value $R_{Kron,min}'=8.0,$ which we found to be the optimal choice in our tests. 
Of course, fine-tuning the parameters might help recovering larger portions of the total flux using \textsc{SExtractor} as well, although we find that the largest impact on this issue is given by the choice of the minimum value allowed to the Kron radius (see next subsection).

The right panel shows the same quantity, but for the average of ten replicas (with different noise realizations) of the LRI simulation. In this case, the total fluxes are evaluated in three ways: (i) using Kron apertures computed on the images as total flux estimation; (ii) using color corrections given by circular apertures, that is, $f_{tot,LRI} = f_{input,HRI} \times f_{ap,LRI}/f_{ap,HRI}$, where $f_{ap,HRI}$ has been computed on a smoothed HRI image, PSF-matched with the LRI one (this is the standard approach used, e.g., for the CANDELS survey); we used two  apertures with diameters of 2 and 3 LRI FWHMs; and (iii) with the same concept as in (ii), but using optimal S/N apertures rather than fixed circular ones.
Here, while all methods yield reasonable results (median $df$ below 5\% for the bulk of the magnitude bins), the offset of the Kron flux measurement is smaller than that of the others; however, the dispersion is larger, as the extension of the error bars indicates in the upper right panel. This is likely due to the increasing contribution of the noise in the pixel summation, and to contamination from neighboring sources. To cope with this issue, a different approach is required, for example, with template-fitting techniques using \textsc{t-phot} or similar software.



\subsubsection{Optimal parameters for a Euclid-like survey} \label{opteuc}

It is interesting to explore in more detail how the accuracy of the flux measurements varies as a function of the \textsc{a-phot} input parameters, namely (i) $R_{Kron,min}'$ and $R_{Kron,max}'$, that is, the minimum and maximum allowed values for $R_{Kron}'$, (ii) $r_{buf}$ and $fac_{rbkgd}$, that is, the inner and outer radius of the circle on which the local background is estimated, and (iii) $q_1$ and  $q_2$, the thresholds adopted in the $\sigma$-clipping procedure for background subtraction. To this aim, we again ran \textsc{a-phot} iteratively on the same dataset, changing these free input parameters. 

The results of the test are shown in Fig. \ref{params}, where the median fractional error $df$ as a function of the input magnitude is shown again. 
The solid lines correspond to runs on the original simulated image, whereas dashed lines show runs on a modified image in which an artificial, constant background of 0.1 counts/sec per pixel has been added. Table \ref{tab1} summarizes the parameters used in each run:
\begin{itemize}
\item the run labeled \texttt{var0} (solid black) has the "optimized" parameter set: $R_{Kron,min}'=8.0$ and $R_{Kron,max}'=20.0$; 
\item \texttt{var1} (solid green) has $R_{Kron,min}'=5.0$ and $R_{Kron,max}'=10.0$;
\item \texttt{var2} (solid red) has $R_{Kron,min}'=3.5$ (the value set as standard in  \textsc{SExtractor}) and $R_{Kron,max}'=10.0$;
\item \texttt{var3} (solid cyan) has $R_{Kron,max}'=20.0$, and all the other parameters are unchanged: the impact of this change is almost negligible except for very faint sources, and indeed the line is almost coincident with the \texttt{var2} line;
\item \texttt{var4} has again the optimized values for $R_{Kron,min}'$ and $R_{Kron,max}'$ but now the image has a background: the accuracy is far from good if \textsc{a-phot} is not allowed to compute the background (dashed magenta), but returns to reasonable results when the option is switched on (\texttt{var5}, dashed yellow, with parameters $q_1=1.5$, $q_2=3.0$, $r_{buf}=10$ and $fac_{rbkgd}=1.2$);
\item finally, setting the background parameters to different values, the accuracy decreases again (\texttt{var6}, dashed black with $q_1=1.0$, $q_2=2.0$, $r_{buf}=5$ and $fac_{rbkgd}=1.5$).
\end{itemize}

The conclusion of this analysis is that it is possible to obtain very accurate results with an adequate tuning of the input parameters. Unfortunately, it is not always possible to explore the parameter space and check the accuracy against some idealized synthetic dataset, or to know a priori the optimal choices in any possible case. Nonetheless, this experiment can shed light on some basic principles. It is worth pointing out that changing the parameters in other software tools like \textsc{SExtractor} depending on the considered dataset would certainly yield a similarly improved accuracy (although it must be taken into account that changes in the \textsc{SExtractor} configuration parameters may often affect detection and deblending as well).

\begin{table}
\caption{\textsc{a-phot} input parameters used in the tests on the simulated Euclid VIS image.}
\centering
\begin{tabular}{| l || l | l | p{4cm} |}
\hline
Run &  $R_{Kron,min}'$ & $R_{Kron,max}'$ &  Background \\ \hline\hline
\texttt{var0} & 8.0 & 20.0 & No \\ \hline
\texttt{var1} & 5.0 & 10.0 & No \\ \hline
\texttt{var2} & 3.5 & 10.0 & No \\ \hline
\texttt{var3} & 3.5 & 20.0 & No \\ \hline
\texttt{var4} & 8.0 & 20.0 & Yes (not measured) \\ \hline
\texttt{var5} & 8.0 & 20.0 & Yes ($q_1=1.5$, $q_2=3.0$, $r_{buf}=10$, $fac_{rbkgd}=1.2$) \\ \hline
\texttt{var6} & 8.0 & 20.0 & Yes ($q_1=1.0$, $q_2=2.0$, $r_{buf}=5$ and $fac_{rbkgd}=1.5$) \\ \hline
\end{tabular}
\end{table} \label{tab1}

\subsection{Background subtraction} \label{bkgtest}

Figure \ref{bkgf} shows the results of a more demanding test on the accuracy of the background estimate by \textsc{a-phot}. We used a crop of the simulated HRI image in two different versions: (i) the original, and (ii) a replica in which an artificial constant background plus additional, local Gaussian-shaped variations have been added. We then ran \textsc{a-phot} on the three images, first without the background estimate, and then switching on the option (also exploring the parameter space before settling to the best choice to which the plots are referred). 

The top panel shows the results of the test on the original image, displaying the histogram of the $df$ quantity described in the previous subsections, as obtained from \textsc{a-phot} with and without the background estimate option, and from a \textsc{SExtractor} run. The three cases yield consistent results, as expected; in particular, no systematic is added by switching on the option in \textsc{a-phot} on perfectly background-subtracted images. The middle panel shows the same quantities, but this time concerning runs on the image (ii) where the background is added: of course, the \textsc{a-phot} run without background subtraction yields incorrect results (out of scale), while the run with the option switched on is still consistent with the \textsc{SExtractor} results.

\subsection{Real data: CANDELS GOODS-South} \label{GS}

\begin{figure}[t!] 
  \includegraphics[width=6cm]{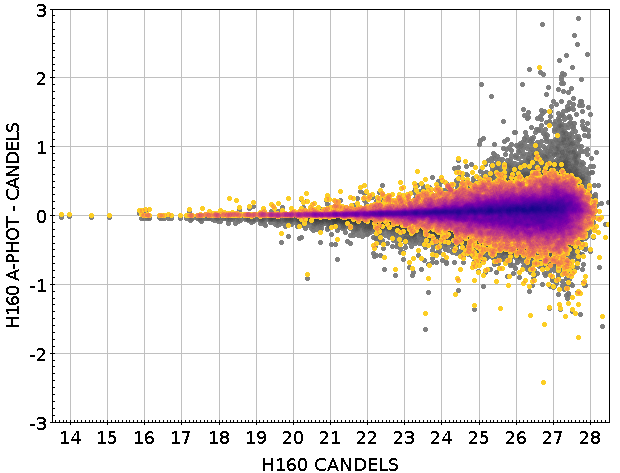}
  \includegraphics[width=6cm]{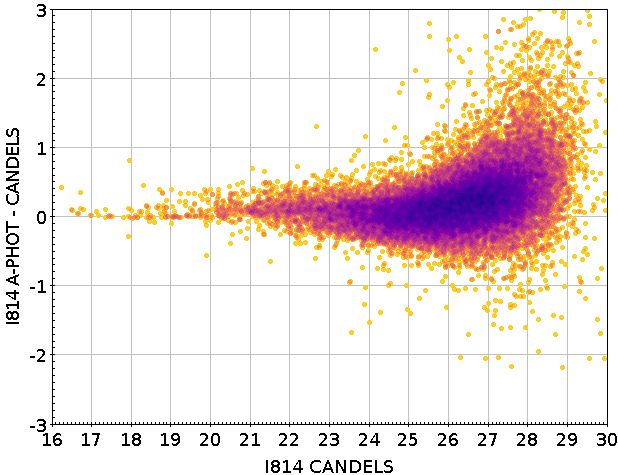}
  \includegraphics[width=6cm]{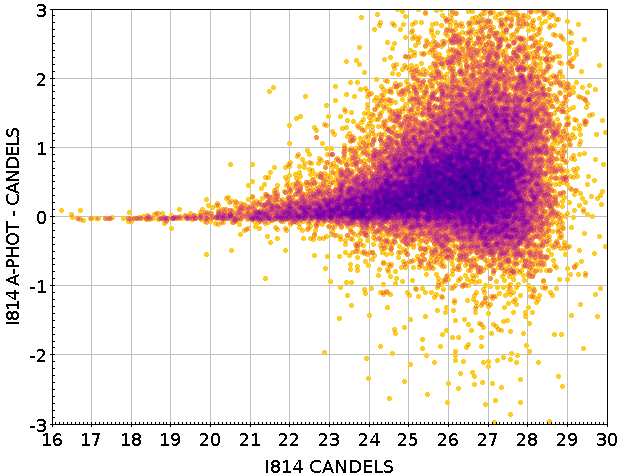}
  \centering
\caption{Performance of \textsc{a-phot} on two GOODS-South CANDELS bands, $H160$ and $I814$, compared to the official CANDELS values obtained with \textsc{SExtractor} (we plot AB magnitudes). Top to bottom: $H160$ with \textsc{a-phot} Kron fluxes, $I814$ with \textsc{a-phot} color-corrected fluxes on optimal S/N apertures, and $I814$ with \textsc{a-phot} Kron flux. The color shading is a proxy for the density of points. In all the cases, we set  $R_{Kron,min}=3.5$; in the top panel, the gray points show the result setting $R_{Kron,min}=8.0$. See text for details.}\label{gs1}
\end{figure}

Finally, we used \textsc{a-phot} to measure the fluxes in the GOODS-South field CANDELS images, and we compared them to those in the official catalog released by the team \citep{Guo2013}. We considered two filters, WFC3 $H160$ and ACS $I814$, and used the CANDELS list of sources and coordinates, excluding those in the ultra-deep field (HUDF) where some subtle issues with the flattening and background subtraction might be present due to the difference of depth with respect to the wide field.

On the $H$ band we used \textsc{a-phot} to compute the total (Kron) flux, internally estimating the morphological parameters. For the $I$ band we used two methods to compute the total flux: (i) the technique described in Section \ref{sims}, that is, $f_{tot,I}=f_{tot,H}\times f_{ap,I}/f_{ap,H}$, where we used a smoothed version $I$-band image, PSF-matched with the $H$-band image, and $f_{ap}$ were measured in "best S/N" elliptical apertures computed on-the-fly on the $H$ band - and the same apertures were used on the $I$ smoothed images; and (ii) an independent Kron measurement on the original $I$ band. In all these cases we used the "optimal" parameters described at the end of Section \ref{sims}, except for the $R_{Kron,min}'$ , which was set to 3.5 rather than 8.0 for consistency with the CANDELS \textsc{SExtractor} estimates.

Figure \ref{apertures} shows a region of the field in which the Kron apertures (with $R_{Kron,min}'=3.5$) and the optimal S/N apertures have been superimposed to the objects. Figure \ref{histGS} shows a histogram of the ratios between the major semi-axis of the two apertures for all the galaxies in the field: compared to the results of a similar test on a simulated dataset (Section \ref{optsn}), the consistency is good, although because we lack an a priori bulge-to-total ratio classification, we obtain a single peak at $R/R_{Kron}' \simeq 0.4$, rather than a bimodal distribution with two peaks (compare with Fig. \ref{sn}).

In Fig. \ref{gs1} we compare the results on the two bands to the CANDELS values, plotting the measured magnitudes. While there is no systematic offset, a scatter at the faint end is present, in particular in the $I814$ band. This can be due to many reasons, including the different way in which the magnitudes are computed (color correction on optimal S/N elliptical apertures and "Kron" elliptical magnitudes rather than color correction on fixed circular apertures in the second and third panel respectively) and the local background subtraction (not used in the \textsc{a-phot} runs, while it is basically unavoidable in \textsc{SExtractor}). The global consistency is nevertheless good, as expected. 

We note that in the $H160$ band a larger scatter toward lower brightnesses would be present taking $R_{Kron,min}'=8.0$ as in the "optimal" Euclid VIS-like case (gray points in the first panel). This is due to the inclusion in the flux summations of pixels with strongly negative values, which remain outside of the aperture when its minimum value is set to 3.5$R_{Kron}$. The difference can be mitigated by switching on the clipping procedure, as described in Section \ref{flux} (not shown).

This test ensures that \textsc{a-phot} yields reasonable results on real datasets. Clearly, a thorough analysis should be performed in order to understand the causes of the discrepancies from case to case; however, this is beyond the scope of this paper.

\section{Summary and conclusions} \label{conclusions}

We presented \textsc{a-phot}, a new code that was specifically designed and developed to perform aperture photometry on astronomical images (typically extragalactic wide  or deep fields) in a straightforward, user-friendly way. 
The rationale behind the development of this software was to obtain a tool that is versatile enough to give accurate results on various cases of astronomical problems, but simple enough to guarantee the full control on the process (which is not always the case with multipurpose, highly engineered tools like \textsc{SExtractor}). At the same time, \textsc{a-phot} includes some interesting new features such as the possibility to perform forced photometry on any list of a priori known positions from any combinations of detection catalogs, and to simultaneously obtain multiple measurements within any number of chosen user-defined circular and elliptical apertures; the on-the-fly determination of optimal elliptical apertures in which the S/N is maximized, an effective on/off switching of the local background estimation and subtraction. These features represent an important step forward toward the accuracy and reproducibility of the measurements and also allow for detailed comparative studies to assess which combination of measurements (flux within Kron aperture, color-corrected fluxes within circular or elliptical apertures, etc.) can be considered the most robust to obtain reliable magnitudes and/or colors in each case of interest.

Given its versatility and accuracy, \textsc{a-phot} can be considered a tool of choice for performing aperture photometry and morphological analyses on upcoming large extragalactic surveys. The software is publicly available and can be downloaded from the website \textit{http://www.astrodeep.eu/a-phot/}.

\section*{Acknowledgements}
The authors acknowledge the contribution of the FP7 SPACE project ASTRODEEP (Ref
. No: 312725), supported by the European Commission.

We thank the anonymous referee for the useful comments and suggestions that greatly helped us to improve the paper.

\bibliographystyle{aa}
\bibliography{mnemonic,biblio}

\end{document}